\documentstyle[12pt]{article}
\begin{document}

\title{Supersymmetric Solutions to the Strong CP Problem}

\author{{\bf S.M. Barr}\\ Bartol Research Institute\\
University of Delaware\\ Newark, DE 19716}

\date{}
\maketitle

\begin{abstract}

It is argued that in the context of supersymmetry, the Strong CP
Problem is most naturally seen as an aspect (particularly severe)
of the whole complex of flavor-violating and CP-violating
problems of supersymmetry. It is shown that certain approaches
to solving these flavor problems also allow simple solutions
to the Strong CP Problem. The idea of ``flavor alignment"
suggested by Nir and Seiberg allows not only flavor violation
to be controlled but supersymmetric contributions to the
theta parameter to be made acceptably small. Another approach
to the flavor-violation problem, namely low-energy supersymmetry
breaking, allows another class of solutions to the Strong CP
Problem to be viable.

\end{abstract}

\newpage

\section{Introduction}

An alternative to the Peccei-Quinn$^1$ solution to the
Strong CP Problem is the idea that $\overline{\theta}$
is small due to an approximate or spontaneously broken
CP invariance of the Lagrangian.$^{2-5}$ There are
a number of facts that commend this approach.

On the negative side is the fact that axions have not
yet been observed, and that experimental constraints
leave only a relatively narrow window in parameter
space where the axion might live. Moreover, the axion
solution requires the existence of a continuous global
symmetry that is exact (except for the QCD anomaly) to
a remarkably high degree. This is problematic from the
point of view of both quantum gravity and superstrings.$^6$

On the positive side there are facts which lend credibility
to the idea that
CP may be a spontaneously broken symmetry. First, theories with few
parameters sometimes conserve CP automatically. 
This appears to be the case with superstrings, where it
has been argued that four-dimensional CP invariance
is actually a local symmetry.$^7$ Second, in theories with
low energy supersymmetry, the soft supersymmetry-breaking terms
have to approximately conserve CP in order not to give
excessive electric dipole moments. 

Indeed, in supersymmetry there is a whole cluster of problems
related to the smallness of flavor-changing and CP-violating
effects that are unsolved.$^8$ It would seem natural to regard
the Strong CP Problem as being among them and to seek a
common approach$^9$ to all these problems, rather than treating
the Strong CP Problem in isolation as the Peccei-Quinn approach 
does.

The attractive idea that a (spontaneously broken) CP
invariance is responsible for the smallness of $\overline{\theta}$
can be implemented in two kinds of models, which we shall call
Type I and Type II. In Type I models the quark mass matrix has
large CP-violating phases in it, which give rise to a Kobayashi-Maskawa
phase, but has a determinant that
is real at tree level because of some flavor symmetry.$^{2,3,4}$
In Type II models the quark mass matrix itself is real at tree 
level,$^5$ so that the CP violation seen in the Kaon system must
be accounted for by some milliweak or superweak force. 

In this paper we are interested in models of Type I. The first
models of this type$^2$ generally had CP invariance broken at the weak
scale by the relative phase of the VEVs of two or more $SU(2)$-doublet Higgs.
It is easy to arrange by some flavor symmetry a pattern of
Yukawa couplings so that this phase appears in the quark mass
matrix, but not (at tree level) in its determinant. An example
is the following triangular form for the up and down quark mass 
matrices, which can be the result of a simple family symmetry:

\begin{equation}
M_d = \left( \begin{array}{ccc}
\lambda_{11} v_0 & \lambda_{12} v_1 & \lambda_{13} v_2 \\
0 & \lambda_{22} v_0 & \lambda_{23} v_1 \\
0 & 0 & \lambda_{33} v_0 
\end{array} \right), M_u = \left( \begin{array}{ccc}
\lambda'_{11} v_0^* & 0 & 0 \\
\lambda'_{21} v_1^* & \lambda'_{22} v_0^* & 0 \\
\lambda'_{31} v_2^* & \lambda'_{32} v_1^* & \lambda'_{33} v_0^*
\end{array} \right).
\end{equation}
 
\noindent
Note that the determinant of the full quark mass matrix
diag$(M_d, M_u)$ only
depends on $\left| v_0 \right|^2$ and does not see the
relative phases of $v_0$, $v_1$, and $v_2$. Since, by the
assumed CP invariance of the lagrangian, the $\lambda_{ij}$
and $\lambda'_{ij}$ are real, $\overline{\theta} = 0$ at tree level.
However, the relative phases of the $v_n$ does lead to a 
non-trivial KM phase, $\delta$.

There are two problems with this type of model. The first, is
that there is
more than one doublet contributing to $M_d$ (or $M_u$) which leads to 
flavor-changing processes mediated by scalar exchange.$^{10}$
The second is that CP is broken spontaneously at the weak scale,
which leads to unacceptable cosmological domain walls. 

Another approach was suggested by A. Nelson in reference 3 and 
further developed in Ref. 4. In this class of models, which we 
will call Nelson models,
there are, in addition to the three families of 
quarks, a vectorlike set of quarks and mirror 
quarks 
that have superlarge $SU(2)_L$-singlet masses. These mirror
quarks mix
with the three families through superlarge
and complex VEVs, which also break CP
spontaneously. Because these complex VEVs appear
in the off-diagonal block which couples the usual families
to the vectorlike quarks,
the determinant of the complete quark mass matrix
(including both light and superheavy states) remains real, but
when the superheavy quarks are integrated out the resulting
light-quark mass matrices have a KM phase. 

This kind of model has neither of the problems that characterized
the earlier models. There is only one Higgs doublet (or in
supersymmetric versions only $H_u$ and $H_d$), and CP
is broken at superlarge scales, so that domain walls can be
inflated away. This approach can also be implemented in
supersymmetry.$^{9,11}$ However, it was pointed out by Dine, Leigh,
and Kagan$^{12}$ that unless ``universality" of the soft 
supersymmetry-breaking terms is satisfied to a high degree of exactness ---
that is, unless the squark masses are highly degenerate and the 
``A terms" are very nearly proportional to the Yukawa terms ---
the down-quark mass matrix and the gluino mass will pick 
up unacceptably large phases
at one loop from the diagrams shown in Fig. 1. But, as 
emphasized by the same authors, there is no {\it a priori} reason
to expect such exact or nearly exact
``universality" to hold. Indeed, in general it would violate the
'tHooft criterion of naturalness, and 
in fact does not seem to hold in superstring models.

The upshot is that the same non-universality that is at the root
of all the well-known flavor and CP problems of supersymmetry
also creates a problem for the $\overline{\theta}$ parameter.
In this sense, it is natural to regard the Strong CP Problem
as another (and particularly severe) aspect
of the general problems of flavor changing and CP violation
in supersymmetry$^8$ and to seek a common approach to solving
all of them. 

In this paper we point out that certain approaches to solving
the general flavor-changing problems of supersymmetry proposed in
recent years also allow one to construct acceptable Type I models
for solving the strong CP Problem.

One possible solution to the flavor-changing problems of supersymmetry,
suggested by Nir and Seiberg$^{13}$, and Nir and Rattazzi$^{14}$
is that the squark and quark mass matrices are ``aligned" because of an
abelian family symmetry. In section 2 we show that this idea allows
acceptable Type I models to be constructed. If the 
fields that spontaneously break the family symmetry also
break CP, the CP violation in the squark and quark mass
matrices can also be aligned in such a way that it cancels in the lowest-order 
contributions to $\overline{\theta}$. These models have a strong
similarity to the older Type I models of Ref. 2, except that
they have a minimal number of Higgs doublets and have CP broken at
large scales as in the Nelson models. In section 3, we observe that
Nelson models avoid the problems pointed out by Dine, Leigh, and Kagan
if supersymmetry is broken at low scales.

\vspace{1cm}

\section{Flavor Alignment Models}

The flavor-alignment approach to the Strong CP Problem in
supersymmetry is similar in spirit to the type of model$^2$
we illustrated in Eq. 1. Indeed, we will consider the following
toy model for simplicity: the effective down and up quark mass
matrices have the forms

\begin{equation}
M_d = \frac{1}{M} \left( \begin{array}{ccc}
\lambda_{11} v' \langle S_0 \rangle & \lambda_{12} v' \langle
S_1 \rangle & \lambda_{13} v' \langle S_2 \rangle \\
0 & \lambda_{22} v' \langle S_0 \rangle & \lambda_{23} v'
\langle S_1 \rangle \\ 0 & 0 & \lambda_{33} v' \langle S_0 \rangle 
\end{array} \right),
\end{equation}

\noindent
and 

\begin{equation}
M_u = \frac{1}{M} \left( \begin{array}{ccc}
\lambda'_{11} v \langle \overline{S}_0 \rangle & 0 & 0 \\
\lambda'_{21} v \langle \overline{S}_1 \rangle & \lambda'_{22} v \langle 
\overline{S}_0 \rangle & 0 \\ 
\lambda'_{31} v \langle \overline{S}_2 \rangle & 
 \lambda'_{32} v
\langle \overline{S}_1 \rangle & \lambda'_{33} v \langle 
\overline{S}_0 \rangle 
\end{array} \right),
\end{equation}

\noindent
where $S_0$, $S_1$, and $S_2$ (and their barred counterparts)
are singlets under the 
Standard Model gauge group, but carry some abelian family 
quantum number. The VEVs of the $S_n$ are roughly of the
same order as the scale $M$, assumed to be large
compared to the weak scale, that appears in the expressions
for $M_d$ and $M_u$. The family symmetry is responsible for the
triangular form of the mass matrices. These singlet scalars,
$S_n$, not only break the family symmetry, but are assumed
to break CP spontaneously as well because of a non-trivial
relative phase among their VEVs.

As in the model descibed in Eq. 1, the determinant of
the full quark mass matrix diag$(M_d, M_u)$ 
does not ``see", at tree level, 
the relative phases of the $\langle S_n \rangle$, so that
$\overline{\theta}_{{\rm tree}} = 0$, whereas a non-trivial
KM phase does result. But, unlike the model of Eq. 1, there is
only one doublet contributing to $M_d$, and one to $M_u$, so that
that FCNC from Higgs exchange is avoided.$^{10}$ Moreover, CP is
broken not by the doublets at the Weak scale but by the singlets
at a very high scale, so that the resulting cosmological domain
walls may be inflated away. One sees, then, that this 
kind of model combines certain features of the older models of
Ref. 2 and the Nelson models of Ref. 3 and 4. We shall see shortly how 
this kind of model solves the problems posed by the diagrams in
Fig. 1 that were pointed out by Dine, Leigh, and Kagan,
but first we must go into more detail.

The matrices given in Eqs. 2 and 3 involve
non-renormalizable operators. These are conceived to
arise from integrating out heavy vectorlike states in a way
now to be described.

Consider that in addition to the ordinary
down-type quarks, $d_n$, and $d^c_n$, $n = 1,2,3$, a vectorlike
set of down-quarks, $D_n$ and $D^c_n$, $n = 1,2,3$ exists. These
new quarks are singlets under $SU(2)_L$. In addition, let there be
a $U(1)_H \times Z_3$ family symmetry. The first, second, and
third generation quarks (both of the $d$ and of the $D$) have
$U(1)_H$ charges $-1$, $0$, and $+1$ respectively. The $d^c_n$
all pick up a phase $e^{2 i \pi/3}$ under the $Z_3$ symmetry, 
while the other down quarks are invariant under it.
The down quark masses come from the following set of Higgs:
an $H_d$ which is an $SU(2)_L$ doublet and neutral under
the family symmetries; a singlet $T$, which gets a superlarge
VEV and is neutral under both the standard model gauge interactions
and the family symmetries; and three singlets, $S_0$, $S_1$, $S_2$,
which have charges $0$, $1$, and $2$ under $U(1)_H$ and pick up
phase $e^{-2 i \pi/3}$ under $Z_3$. The VEVs of the $S_n$, which are
assumed to be of the same order (roughly) as that of $T$, not
only break the abelian family symmetries, but their relative
phases are assumed
to spontaneously break CP as well.  

With this set of fields the down-type quarks have a $6 \times 6$
mass matrix of the form

\begin{equation}
(d_m | D_m) \left( \begin{array}{ccc|ccc}
0 & & & \lambda'_1 \langle H_d \rangle & & \\
& 0 & & & \lambda'_2 \langle H_d \rangle & \\
& & 0 & & & \lambda'_3 \langle H_d \rangle \\ \hline
\lambda_{11} \langle S_0 \rangle & \lambda_{12} \langle S_1 \rangle 
& \lambda_{13} \langle S_2 \rangle & \lambda''_1 \langle T \rangle
& & \\ 0 & \lambda_{22} \langle
S_0 \rangle & \lambda_{23} \langle
S_1 \rangle & & \lambda''_2 \langle T \rangle & \\
0 & 0 & \lambda_{33} \langle S_0 \rangle & & & 
\lambda''_3 \langle T \rangle
\end{array} \right) \left( \begin{array}{c}
d^c_n \\ \hline D^c_n
\end{array} \right), 
\end{equation}

\noindent
which we shall abbreviate as

\begin{equation}
(d | D) \left( \begin{array}{c|c}
0 & H \\ \hline S & T \end{array} \right) \left( \begin{array}{c}
d^c \\ \hline D^c \end{array} \right).
\end{equation}

\noindent
There is an entirely analogous structure for up quarks, with in addition
to $u_n$ and $u^c_n$ a set of $U_n$ and $U^c_n$. The $U(1)_H$ charges
are again $-1$, $0$ and $+1$ for the three families, and the $u^c$ are
assumed to pick up a phase $e^{-2 i \pi/3}$ under $Z_3$. There are
in addition to the above named Higgs, an $H_u$, which is a doublet
neutral under family symmetry, and a set of $\overline{S}_n$, $n = 0,1,2$,
whose quantum numbers are opposite to those of $S_n$. The $6 \times 6$
mass matrix of the up quarks will have a form analogous to that in Eq. 4. 
It is clear that upon integrating out the heavy states one is left
with an effective $3 \times 3$ mass matrix for the light down quarks
that is given by the familiar ``see-saw" formula $M_d = - H \; T^{-1} \; S$,
and similarly for the up mass matrix. These have just the forms given in 
Eqs. 2 and 3.

As already stated, CP violation is supposed to arise
spontaneously as a result of non-trivial relative
phases among the VEVs of the $S_n$ and $\overline{S}_n$.
For concreteness, we can imagine that the superpotential
of the Higgs that get superlarge VEVs has the 
$U(1)_H \times Z_3$-invariant form

\begin{equation}
\begin{array}{ccl}
W & = & (T^2 - M_T^2) \; X_T + (S_0 \overline{S}_0 - M_0^2) \;X_0 \\
  & + & (S_1 \overline{S}_1 - M_1^2) \; X_1 + (S_2 \overline{S}_2
+ M_2^2) \; X_2 \\
 & + & (S_0^3 + \overline{S}_0^3) \\
 & + & (S_2 \; \overline{S}_1^2 \; S_0/M^2 + Z^2) \; Y +
(\overline{S}_2 \; S_1^2 \; \overline{S}_0/M^2 + Z^2) \; \overline{Y}.
\end{array}
\end{equation}

\noindent
This is not the only possibility, but is simple and illustrates
the essential idea. One finds upon minimizing the (supersymmetric)
scalar potential that arises from this that 
$\langle T \rangle = M_T$, $\langle S_0 \rangle = \langle \overline{S}_0
\rangle = M_0$, $\langle S_1 \rangle = \langle \overline{S}_1 \rangle
= M_1$, and $\langle S_2 \rangle = \langle \overline{S}_2 \rangle
= i \; M_2$, where the $M$'s are all real
parameters by the CP invariance of the lagrangian. The phase in
the VEV of $S_2$ comes from the plus sign in the term involving $X_2$
in $W$. We shall discuss corrections to these phases later.
Notice also that when the scalar component of $W$ is evaluated
at the minimum it is real, as is necessary$^{11}$ if the A parameters
are to be real at tree level in supergravity theories.

The physical, CP-violating order parameter must be invariant
under the family symmetry $U(1)_H \times Z_3$. The lowest-dimension
such operator is $S_2 \overline{S}_1^2 S_0$ or equivalently
$S_2 S_1^{*2} S_0$. This is the combination that in fact comes
into the expression for the leading contribution to $\overline{\theta}$.
Because it involves (as can be seen from Eq. 2) several small 
flavor-changing Yukawa interactions, these contributions 
will end up being quite suppressed, as will now be shown.

As noted, the mass matrix of Eq. 4 has three superheavy eigenstates and
three light eigenstates, with the effective mass matrix of the three
light states being given by the ``see-saw" formula

\begin{equation}
M_d = - H \; T^{-1} \; S,
\end{equation}

\noindent
where the notation is defined in Eq. 5.
It is convenient to define the combination
$s \equiv T^{-1} S = - H^{-1} M_d$. Note that
$s$, $S$, and $M_d$ all have the same triangular form.
If we assume
for simplicity that all the (non-zero) elements of $H$ 
are of order $\langle H_d \rangle \equiv v'$, then
$s_{mn} \sim (M_d)_{mn}/v'$. Using the Wolfenstein
parameter, $\lambda$, one can then write

\begin{equation}
\begin{array}{l}
s_{33} \sim m_b/v', \\
\\
s_{23} \sim s_{22} \sim \lambda^2 m_b/v', \\
\\
s_{13} \sim s_{12} \sim \lambda^3 m_b/v', \\
\\
s_{11} \sim \lambda^4 m_b/v'.
\end{array}
\end{equation}

\noindent
One is now in a position to estimate the contributions
to $\overline{\theta}$ coming from the dangerous diagrams in Fig. 
1. From what has already been said, it is obvious that
they will be proportional to Im$(s_{13} s_{23}^* s_{22} s_{12}^*)$,
which is of order $\lambda^{10}$. Because of small denominators
the leading contribution to $\overline{\theta}$ will be lower
order in $\lambda$ than this, as we shall see.

The contribution to the down-quark mass matrix coming from
Fig. 1 (a) is proportional to the $A$ parameter and to the
gluino mass, so that one may ignore the part coming from
superheavy states circulating in the loop as these will be
suppressed by $(m_{SUSY}/M_{GUT})^2$. Let us therefore
block-diagonalize the $6 \times 6$ matrix given in Eq. 4,
which we will call ${\cal M}$, to separate the light and
superheavy states. This is done by ${\cal M} \longrightarrow
U_L \; {\cal M} \; U_R^{\dag}$, where the unitary matrices are
given by

\begin{equation}
\begin{array}{l}
U_L \cong \left( \begin{array}{cc}
I & - H (I + s s^{\dag})^{-1} T^{-1} \\
T^{-1} (I + s s^{\dag})^{-1} H & I 
\end{array}
\right), \\
U_R = \left( \begin{array}{cc}
(I + s^{\dag} s)^{-\frac{1}{2}} & 0 \\
0 & (T(I + s s^{\dag})T)^{-\frac{1}{2}}T 
\end{array} \right) \left( \begin{array}{cc}
I & -s^{\dag} \\ s & I \end{array} \right).
\end{array}
\end{equation}

\noindent
This gives, when applied to the matrix in Eq. 5,
$M_d = -H \; s$, as it should. The same transformations applied
to the squark fields will also separate the light and superheavy squarks, since
the superheavy elements of the squark mass matrices are
just given by the SUSY-invariant terms. In the original basis,
the SUSY-breaking $LL$ and $RR$ squark mass terms are 
just diagonal matrices whose elements are all of order
$m_0^2$. They are diagonal because of the abelian family symmetry.
Of course, loop effects and gravity effects can induce family-changing
elements of these matrices, but these will involve insertions of
the VEVs of the $S_n$ divided by either the Planck scale or the
scale of $\langle T \rangle$. Thus the contributions 
to $\overline{\theta}$ so produced are
smaller than the ones we shall consider.

The $LR$ squark mass term has, in the original basis, 
the same form as the mass term given
in Eq. 5, namely 

\begin{equation}
m_{LR}^2 = A_0 \left( \begin{array}{cc}
0 & \tilde{H} \\ \tilde{S} & \tilde{T} 
\end{array} \right).
\end{equation}

\noindent
Here, $\tilde{H}$, $\tilde{S}$, and $\tilde{T}$ are matrices
which have the same structure as $H$, $S$, and $T$, because
of family symmetry, namely $\tilde{H}$ is diagonal and proportional
to $\langle H_d \rangle$, $\tilde{T}$ is diagonal and proportional to
$\langle T \rangle$, and $\tilde{S}$ is triangular and proportional
to $\langle S_n \rangle$. We can define by analogy to $s$ a 
dimensionless matrix $\tilde{s} \equiv \tilde{T}^{-1} \; \tilde{S}$.
In a model in which the inter-family hierarchy was completely
explained by family symmetry, one would expect that $\tilde{s}$ would
exhibit the same hierarchy as $s$, so that $\tilde{s}_{mn}$ would
be of the same order in the Wolfenstein parameter as $s_{mn}$.
In the toy model we are discussing, there is not enough family 
symmetry to completely explain the hierarchy. The smallness of
the non-vanishing off-diagonal elements of $s$ can be explained
by supposing that $\langle S_1 \rangle$ and $\langle S_2 \rangle$ 
are small compared to $\langle S_0 \rangle $. Then one expects that 
the non-vanishing
off-diagonal elements of $\tilde{s}$ are also small and of the same
order in $\lambda$. However, in this toy model the hierarchy
among the diagonal elements of $s$ is not explained, since they
all come from the VEV of $S_0$. No symmetry principle thus demands
that the diagonal elements of $\tilde{s}$ must have a similar
hierarchy. Actually, this will not turn out to matter, since
even if only the off-diagonal elements of $\tilde{s}$ are
suppressed, a sufficient suppression of $\overline{\theta}$ will 
result. But we will assume, since it is simpler, and since one 
expects it to be true in a theory where family symmetry explains the
fermion mass hierarchy, that $\tilde{s}_{mn} \sim s_{mn}$.

After applying the transformations $U_L$ and $U_R$ to the
squark-mass matrices one finds that they have the forms

\begin{equation}
\begin{array}{ccl}
m^2_{LL,{\rm light}} & = & m^2_{LL(d)}, \\
\\
m^2_{RR,{\rm light}} & = & (I + s^{\dag}s)^{-\frac{1}{2}} m^2_{RR(d^c)}
(I + s^{\dag}s)^{-\frac{1}{2}} \\
& + &  (I + s^{\dag}s)^{-\frac{1}{2}} s^{\dag} m^2_{RR(D^c)} s 
(I + s^{\dag}s)^{-\frac{1}{2}} \\
\\
m^2_{LR,{\rm light}} & = & (- \tilde{H} s - H \; 
(I + s s^{\dag})^{-1} T^{-1} \tilde{T} [\tilde{s} - s])
(I+s^{\dag}s)^{-\frac{1}{2}}.
\end{array}
\end{equation}

\noindent
If we take the lowest
order in the small quantities $s$ and $\tilde{s}$, $m^2_{LL,{\rm light}}$
and $m^2_{RR,{\rm light}}$ are diagonal matrices, and the
matrix $m^2_{LR,{\rm light}}$ is equal to $(-\tilde{H}s
-HT^{-1}\tilde{T}[\tilde{s} - s])$, which has the same triangular
form as $s$ and $\tilde{s}$. Moreover, in the supersymmetric basis in 
which we are working, the gluino couplings are flavor-diagonal.
Thus, at lowest order in $s$, 
the result of doing the loop in Fig. 1 (a) is just to give
a contribution to the light down-quark mass matrix which (like the
tree level term) has the same triangular form as $s$. Hence, the 
determinant of the quark mass matrix is still real at one loop
if we ignore terms of higher order 
in $s$ and $\tilde{s}$. Of course, this trivially had to be the case, 
since the
CP-violating invariant, as noted above, is $s_{31} s_{32}^* s_{22}
s_{21}^*$. But if one now keeps higher order terms in $s$ and $\tilde{s}$,
one finds a number of contributions to $\overline{\theta}$ all of 
which are of order

\begin{equation}
\delta \overline{\theta}  \sim  \frac{8}{3} \frac{\alpha_s}{4 \pi} 
\frac{A_0 m_g}
{m_0^2} \left( \frac{s_{21} s_{31} s_{32}}{s_{22}} \right) 
\sim 2 \times 10^{-2} \lambda^6 \left( \frac{A_0 m_g}{m^2_0} \right)
\left( \frac{m_b}{v'} \right)^2,
\end{equation}

\noindent
or

\begin{equation} 
\overline{\theta} \sim  10^{-9} \left( \frac{A_0 m_g}{m^2_0} \right) 
\tan^2 \beta.
\end{equation}

\noindent
We have assumed that the phases in $s$ are of order unity, as is
necessary if the KM phase is to be of order unity.

The contribution to $\overline{\theta}$ that comes from
Fig. 1 (b) is smaller than that just given. It is typically
of order $\frac{\alpha_s}{4 \pi} \frac{A_0}{m_g} {\rm Im}
(s_{31} s_{32}^* s_{22} s_{21}^*) \sim \lambda^{10}$. 

In addition, one must investigate the phase of $\langle S_0 \rangle$ 
and $\langle \overline{S}_0 \rangle$. First, there are
possible gravitational effects. For example, there can be
a term in the superpotential for the $S_n$ that has the form
$\epsilon_G S_2 \overline{S}_1^2 S_0 X_0/M_{Pl}^2$, where $\epsilon_G$
is some suppression factor that depends on the nature of
the Planck-scale physics and is presently uncalculable. The $F_{X_0} = 0$
equation then gives a phase to $\langle S_0 \rangle$ that is
of order $\epsilon_G \langle S_n \rangle^2/M_{Pl}^2$. 
The phase of $\langle S_0 \rangle$ contributes
directly at tree level to $\overline{\theta}$. Thus one
requires that the family symmetries and CP are broken
at scales less than or of order $10^{14.5} {\rm GeV}
\epsilon_G^{-\frac{1}{2}}$. This is perfectly consistent
with a GUT-scale breaking, especially since 
$\epsilon_G$ may be small.

Another contribution to the phase of $\langle S_0 \rangle$
comes from supersymmetry breaking. In particular, there will
be a supersymmetry-breaking term of the form 
$A (S_2 \overline{S}_1^2 S_0 Y)/M^2$ in the scalar
potential for $S_0$ in addition to the term $\left|
S_0^2 - M_0^2 \right|^2$. In the supersymmetric limit $Y$ has
vanishing VEV, so that this term will not matter. However, 
when supersymmetry is broken $Y$ gets a VEV of order $A$.
Therefore there will be induced a linear term for $S_0$ that
contains a CP-violating phase. The contribution to the phase
of $\langle \overline{S}_0 \rangle \langle S_0 \rangle$ 
(which appears in the determinant
of the full quark mass matrix and thus in $\overline{\theta}$) is
of order $m_{SUSY}^2/M^2$, which is easily smaller than 
$10^{-9}$.

The model we have presented above is far from being
unique. The triangular form has been chosen for
simplicity, but there are many other forms that would
admit the same kind of solution of the strong CP problem.
For example, by having the non-zero elements of the
$s$ matrix be the $11$, $22$, $33$, $12$, $13$, and $32$,
with the phase being in the $13$ element, the leading
contribution to $\overline{\theta}$ is suppressed by
order $\lambda^8$. There are patterns that would
allow a suppression by $\lambda^{10}$ but these seem less
realistic. (For example, having the non-zero elements being
$11$, $33$, $31$, $13$, $32$, and $23$.) In any event,
it would seem to be generally the case that $\overline{\theta}$
in this kind of model should be not far below $10^{-11}$.

A challenge for any completely satisfactory theory of
flavor is to find a family symmetry that aligns
the quarks and squarks in such a way that both the
Strong CP Problem and the other flavor problems
of supersymmetry are solved at the same time. 

\section{Nelson Models and Low-energy Supersymmetry Breaking}

The basic idea of the models proposed in Refs. 4 and 5
is very simple. Let $d_i + d^c_i$ be the usual three
families of down quark, and $D_I + D^c_I$ be a set of
mirror down quarks which are $SU(2)_L$ singlets. (Such
a set of particles can arise in $SU(5)$, for example from
having three sets of ${\bf 10} + \overline{{\bf 5}}$ and
some additional sets of ${\bf 5} + \overline{{\bf 5}}$.)
The Yukawa interactions of the down quarks 
are assumed to have the following form

\begin{equation}
{\cal L}_{down \; mass} = (d_i, \; D_I) 
\left( \begin{array}{cc}
\lambda_{ij} \langle H \rangle & 0 \\
\lambda^m_{Ij} \langle S_m \rangle & \lambda_{IJ} \langle T
\rangle
\end{array} \right) \left( \begin{array}{c}
d^c_j \\ D^c_J \end{array} \right).
\end{equation}

\noindent
$H$ is just the Higgs doublet of the Standard Model.
(For the moment we are describing a non-supersymmetric model.)
Its vacuum expectation value does not have a physically
meaningful phase. (It can be changed by a global weak hypercharge
rotation.) $S_m$ and $T$ are singlets with VEVs of the same order
and much larger than the Weak scale. (It is natural but not
necessary to take this to be near the unification scale.)
$\langle T \rangle$ is assumed not to break CP , while
$\langle S_m \rangle$ have relative phases that do violate CP
spontaneously. By the CP invariance of the lagrangian,
the Yukawa couplings $\lambda$ are real. Here it should be
noted, in contrast to the models discussed in the last section,
no family symmetry need distinguish among the $d_i$ or among
the $S_n$, but some symmetry does distinguish the $d$ from the $D$.

The up quark mass matrix has the simple $3 \times 3$ form
$u_i u^c_j \langle H^* \rangle$. Thus it is easy to see that
the complete quark mass matrix, including the heavy states, has
determinant proportional to $\left| \langle H \rangle \right|^6
\langle T \rangle^n$ (where there are $n$ mirror quarks). Since
this is real, and $\theta_{QCD}$ is real by the CP invariance
of the Lagrangian, $\overline{\theta}$ vanishes at tree level.
On the other hand, when the superheavy states are integrated out,
one is left with three families of light quarks, which have a non-trivial
KM phase (coming from the VEVs of the $S_m$). The model is
thus indistinguishable from the KM model at low energy, but has
no Strong CP Problem.$^{3,4}$

The severe difficulty pointed out by Dine, Leigh, and Kagan$^{12}$ 
is that the diagrams in Fig. 1 are murderous here if there
are order unity violations of ``universality" in the 
supersymmetry-breaking terms. In fact, $\overline{\theta}$
is expected to be of order $\alpha_s/4\pi$ in that case.

This is not the case, however, if SUSY is broken at low scales.$^{15}$
For simplicity, consider an $SU(5)$ model where supersymmetry
breaking is communicated from some hidden sector by a singlet
field, $S$, to a ``messenger sector" consisting of a ${\bf 5} +
\overline{{\bf 5}}$, which have masses of order $100$TeV. 
From the messenger sector, the supersymmetry
breaking is communicated to the known particles and their
superpartners by $SU(3) \times SU(2) \times U(1)$ gauge interactions.
In addition, let there be, as discussed above, a set of mirror
fields, $D_I + D^c_I$, that are contained in some other
set of ${\bf 5} + \overline{{\bf 5}}$ representations, denoted
${\bf 5}_I + \overline{{\bf 5}}_I$. These are assumed
to have superlarge masses. That the down-quark 
mass matrix have the form given in Eq. 14, in particular that
there be no $d_i D^c_J H_d$ coupling, requires that some
symmetry distinguish the $D^c_I$ from the $d^c_i$. (They have
the same $SU(3) \times SU(2) \times U(1)$ quantum numbers,
however.)

It is easy to see that in such a model, the diagrams in Figs. 1
and 2 are not dangerous. Simply integrating out the superheavy
mirror quarks, ${\bf 5}_I + \overline{{\bf 5}}_I$, 
leads to a low energy theory that is
nothing but the MSSM together with the low-energy SUSY breaking
sector and messenger sector. This effective low-energy theory
has a KM phase (from the phases of $\langle S_m \rangle$ as discussed
above), but no tree-level $\overline{\theta}$. The non-universality
of the soft SUSY breaking terms is then small enough to render
the one-loop contributions in Fig. 1 harmless.
In particular, since the flavor symmetry of the Standard Model gauge
interactions is broken only by Yukawa interactions,
the non-universality of the soft supersymmetry-breaking terms
involving the ordinary quarks is suppressed by powers of 
$(m_q/\Lambda)^2$,
where $m_q$ is a light quark mass and $\Lambda \sim 100$TeV is the scale
of SUSY breaking.

\section*{References}

\begin{enumerate}

\item R. Peccei and H.R. Quinn, {\it Phys. Rev. Lett.} {\bf 38}, 1440 (1977).
\item M.A.B. Beg and H.S. Tsao, {\it Phys. Rev. Lett.} {\bf 41}, 278 (1978);
H. Georgi, {\it Had. J.} {\bf 1}, 155 (1978); R.N. Mphapatra and 
G. Senjanovic, {\it Phys. Lett.} {\bf 126B}, 283 (1978); G. Segre and 
H.A. Weldon, {\it Phys. Rev. Lett.} {\bf 42}, 1191 (1978);
S.M. Barr and P. Langacker, {\it Phys. Rev. Lett} {\bf 42}, 1654 (1978).
\item A. Nelson, {\it Phys. Lett.} {\bf 136B}, 165 (1984).
\item S.M. Barr, {\it Phys. Rev. Lett.} {\bf 53}, 329 (1984);
{\it Phys. Rev.} {\bf D30}, 1805 (1984).
\item S.M. Barr and D. Seckel, {\it Nucl. Phys.} {\bf B233}, 116 (1984); 
S.M. Barr and A. Zee, {\it Phys. Rev. Lett.} {\bf 55}, 2253 (1985);
A. Dannenberg, L.J. Hall, and L. Randall, {\it Nucl. Phys.} {bf B271},
574 (1986); L.J. Hall and L. Randall, {\it Nucl. Phys.} {\bf B274}, 157
(1986).
\item For discussion of gravitational effects and global symmetries
see T. Banks, {\it Physicalia} {\bf 12}, 19 (1990); L. Krauss and
F. Wilczek, {\it Phys. Rev. Lett.} {\bf 62}, 1221 (1989); A. Strominger
in {\it Proceedings of the Theoretical Advanced Study Institute in
Elementary Particles, Providence R.I., 1988} (ed. A. Jevicki and
C.-I. Tan, World Scientific, Singapore, 1989) and refs. therein.
For discussion of superstrings and global symmetries see
T. Banks, L. Dixon, D. Friedan, and E. Martinec, {\it Nucl. Phys.}
{\bf B299}, 613 (1988).
\item M. Dine, R. Leigh, and D. MacIntyre, {\it Phys. Rev. Lett.}
{\bf 69}, 2030 (1992).  
\item L.J. Hall, A. Kostelecki, and S. Raby, {\it Nucl. Phys.},
{\bf B267}, 415 (1986); H. Georgi, {\it Phys. Lett.} {\bf 169B},
231 (1986); M. Dine, R. Leigh, and A. Kagan, {\it Phys. Rev.} {\bf D48},
4269 (1993).
\item S.M. Barr and G. Segre, {\it Phys. Rev.} {\bf D48}, 302 (1993).
\item S.L. Glashow and S. Weinberg, {\it Phys. Rev.} {\bf D15}, 1958
(1977).
\item S.M. Barr and A. Masiero, {\it Phys. Rev.} {\bf D38}, 366 (1988).
\item M. Dine, R. Leigh, and A. Kagan, {\it Phys. Rev.} {\bf D48},
2214 (1993).
\item Y. Nir and N. Seiberg, {\it Phys. Lett.} {\bf 309B}, 337 (1993).
\item Y. Nir and R. Rattazzi, RU-96-11, WIS-96/11, hep-ph/9603233.
\item See, for example, M. Dine and A. Nelson, {\it Phys. Rev.}
{\bf D48}, 1277 (1993); M. Dine, A. Nelson, and Y. Shirman, {\it
Phys. Rev.} {\bf D51}, 1362 (1995).
\end{enumerate}

\newpage

\noindent
{\bf\large Figure Captions}

\noindent
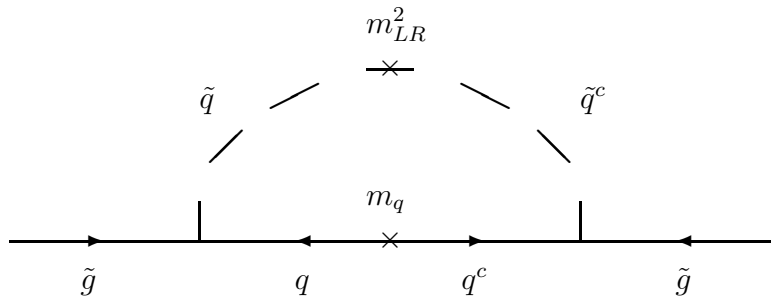
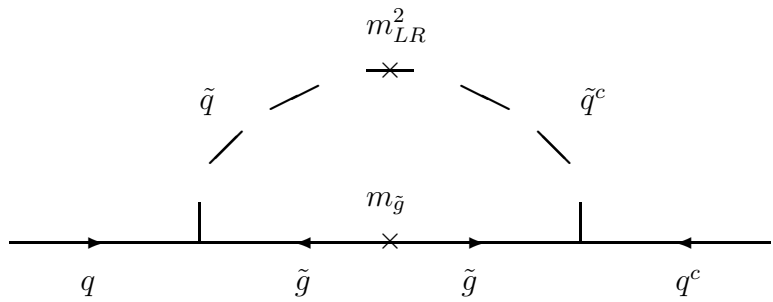
{\bf Fig. 1:} In supersymmetric models where $\overline{\theta}$ vanishes
at tree level due to a spontaneously broken CP invariance,
in general diagram (a) gives too large a phase to the 
mass of the gluino, and diagram (b) gives too large a contribution
to the phase of $\det M_q$.

\newpage

\begin{picture}(360,216)
\thicklines
\put(108,72){\line(0,1){15}}
\put(112,102){\line(1,1){12}}
\put(135,122.5){\line(2,1){18}}
\put(171,137){\line(1,0){18}}
\put(207,131.5){\line(2,-1){18}}
\put(236,114){\line(1,-1){12}}
\put(252,87){\line(0,-1){15}}
\put(36,72){\vector(1,0){36}}
\put(72,72){\line(1,0){72}}
\put(180,72){\vector(-1,0){36}}
\put(180,72){\vector(1,0){36}}
\put(216,72){\line(1,0){72}}
\put(324,72){\vector(-1,0){36}}
\put(175.5,69.5){$\times$}
\put(175.5,134.5){$\times$}
\put(63,54){$\tilde{g}$}
\put(288,54){$\tilde{g}$}
\put(144,54){$q$}
\put(207,54){$q^c$}
\put(171,86){$m_q$}
\put(171,151){$m^2_{LR}$}
\put(108,122.5){$\tilde{q}$}
\put(252,122.5){$\tilde{q}^c$}
\put(160,0){\bf Fig. 1 (a).}
\end{picture}

\begin{picture}(360,216)
\thicklines
\put(108,72){\line(0,1){15}}
\put(112,102){\line(1,1){12}}
\put(135,122.5){\line(2,1){18}}
\put(171,137){\line(1,0){18}}
\put(207,131.5){\line(2,-1){18}}
\put(236,114){\line(1,-1){12}}
\put(252,87){\line(0,-1){15}}
\put(36,72){\vector(1,0){36}}
\put(72,72){\line(1,0){72}}
\put(180,72){\vector(-1,0){36}}
\put(180,72){\vector(1,0){36}}
\put(216,72){\line(1,0){72}}
\put(324,72){\vector(-1,0){36}}
\put(175.5,69.5){$\times$}
\put(175.5,134.5){$\times$}
\put(63,54){$q$}
\put(288,54){$q^c$}
\put(144,54){$\tilde{g}$}
\put(207,54){$\tilde{g}$}
\put(171,86){$m_{\tilde{g}}$}
\put(171,151){$m^2_{LR}$}
\put(108,122.5){$\tilde{q}$}
\put(252,122.5){$\tilde{q}^c$}
\put(160,0){\bf Fig. 1 (b).}
\end{picture}

\end{document}